\documentclass[prd,aps,showpacs,groupedaddress,eqsecnum,notitlepage,nofootinbib,superscriptaddress]{revtex4-2}
\usepackage{graphicx}

\usepackage{tikz,siunitx,mwe}
\usepackage{tensor}
\usepackage{epstopdf}
\usepackage{amsmath}
\usepackage{amsfonts}
\usepackage{slashed}
\usepackage{amssymb}
\usepackage{enumerate, color}
\usepackage{subfigure}
\usepackage{lipsum}
\usepackage{color}
\usepackage{graphicx,bm}
\usepackage[utf8]{inputenc}
\usepackage{eurosym}
\usepackage{scalerel}%To smaller subindex
\usepackage{float}
\usepackage{accents}
\usepackage[colorlinks=true,pdfstartview=FitV,linkcolor=blue,citecolor=blue,urlcolor=blue,breaklinks=true]{hyperref}
\usepackage{orcidlink}
\newcommand{\be}{\begin{equation}}
\newcommand{\ee}{\end{equation}}
\newcommand{\ben}{\begin{eqnarray}}
\newcommand{\een}{\end{eqnarray}}
\newcommand{\bes}{\begin{subequations}}
\newcommand{\ees}{\end{subequations}}
\usepackage{comment}
\def\bal#1\eal{\begin{align}#1\end{align}}

\def\bal#1\eal{\begin{align}#1\end{align}}
\begin{document}
\title{Rotation effects on the graphene wormhole energy levels}

%%%%%%%%%%%%%%%%%%%%%%%%%%%%%%%%%%%%%%%%%%%%%%%%%%%%%%%%%%%%%%%%%%%%%%%%%%%%%%

%-------------------------------------------%
\author{G. Q. Garcia\,\orcidlink{0000-0003-3562-0317}}
\email{gqgarcia99@gmail.com}
\affiliation{Centro de Ci\^encias, Tecnologia e Sa\'ude, Universidade Estadual da Para\'iba, 58233-000, Araruna, PB, Brazil.}
%-------------------------------------------%
%-------------------------------------------%
\author{P.J. Porf\'irio\,\orcidlink{0000-0002-4713-3332}}
\email{pporfirio@fisica.ufpb.br}
\affiliation{Departamento de F\'isica, Universidade Federal da Para\'iba, 58051-970, Jo\~ao Pessoa, PB, Brazil. }
%-------------------------------------------%
%-------------------------------------------%
\author{C. Furtado\,\orcidlink{0000-0002-3455-4285}}
\email{furtado@fisica.ufpb.br}
\affiliation{Departamento de F\'isica, Universidade Federal da Para\'iba, 58051-970, Jo\~ao Pessoa, PB, Brazil. }
%-------------------------------------------%
%-------------------------------------------%
\author{D. C. Moreira\,\orcidlink{0000-0002-8799-3206}}
\email{moreira.dancesar@gmail.com}
\affiliation{Centro de Ci\^encias, Tecnologia e Sa\'ude, Universidade Estadual da Para\'iba, 58233-000, Araruna, PB, Brazil.}
%-------------------------------------------%
%%%%%%%%%%%%%%%%%%%%%%%%%%%%%%%%%%%%%%%%%%%%%%%%%%%%%%%%%%%%%%%%%%%%%%%%%%%%%%
\begin{abstract}
In this work, we are interested in how spinning effects influence the electronic properties of the graphene wormhole. For this purpose, we have described the graphene by the wormhole background based on the model developed by Gonz\'alez and his co-workers. By applying a coordinate transformation in the metric of graphene wormhole, we can introduce rotating effects. In the continuum limit, by solving the massless Dirac equation in the context of a rotating wormhole background, we obtain the Landau levels for the rotating graphene wormhole. We still have exposed the analogy between the graphene wormhole and fermions on the G\"odel-type spacetime.  
\end{abstract}
%%%%%%%%%%%%%%%%%%%%%%%%%%%%%%%%%%%%%%%%%%%%%%%%%%%%%%%%%%%%%%%%%%%%%%%%%%%%%%
\keywords{Graphene, Wormhole Background, Topological Defects, Landau Levels}
\pacs{03.65.Ge, 03.65.Vf}
%%%%%%%%%%%%%%%%%%%%%%%%%%%%%%%%%%%%%%%%%%%%%%%%%%%%%%%%%%%%%%%%%%%%%%%%%%%%%%
\maketitle
%%%%%%%%%%%%%%%%%%%%%%%%%%%%%%%%%%%%%%%%%%%%%%%%%%%%%%%%%%%%%%%%%%%%%%%%%%%%%%
\section{Introduction}

Gravitational analogue models in condensed matter systems have proven to be quite useful as theoretical laboratories for studying several systems (see, \textit{e.g.},  \cite{Barcelo:2005fc} for a review). As an example, the geometric theory of defects in solids allows us to relate the presence of topological defects in materials to some background metric. As a consequence, depending on the system, we can use techniques from quantum field theory on curved spacetimes to deal with various results from table-top experiments \cite{volterra1907equilibre, moraes2000condensed}. From this viewpoint, Katanaev and Volovich showed that disclinations are related to the behavior of the curvature tensor of the geometric background, while dislocations are associated with its torsion tensor \cite{katanaev1992theory, katanaev2023combined}. One can use these ideas to study graphene, which stands out for being a semimetal characterized by a honeycomb-like carbon lattice structure and conical band structures around its Fermi points. Another unusual property of graphene is the presence of topologically protected zero modes whose study has attracted great attention in the last decade due to the versatility of possible applications in various branches of technology \cite{castro2009electronic, bena2009remarks,katsnelson2007graphene}. Among its key properties, its valence band touches the conduction band at its six Fermi points, and in the vicinity of such points,  graphene presents a linear dispersion relation. In this way, the particle dynamics on the graphene sheet can be modeled by the behavior of relativistic quasi-particles described by the Weyl equation in the low energy regime for the tight-biding model description \cite{katsnelson2007graphene}. 

Vozmediano {\it et al.}, studying the behavior of gauge fields in graphene in the presence of topological defects and soft deformations, proved that the formation of pentagonal or heptagonal rings in the graphene lattice is related to disclinations \cite{vozmediano2010gauge}. In contrast, the formation of pentagon-heptagon pairs is linked to dislocations in the lattice \cite{carpio2008dislocations}. Thus, it is possible to build different gravitational analogue models of graphene and several studies have emerged since then \cite{puntigam1997volterra, lammert2004graphene}. For example, the spectrum of the fullerene molecule $C_{60}$ can be described by a graphene lattice folded into a polyhedron in the continuum limit \cite{gonzalez1993electronic, PhysRevLett.69.172} and its cosmic string analogue is assembled from the introduction of pentagon-heptagon pairs and Stone-Wales defects \cite{cortijo2007effects}. In addition, gravitational models equipped with compact extra-dimension such as Kaluza-Klein theory can be used to describe geometric phases in graphene layers with defects \cite{bakke2012kaluza}. 

An interesting gravitational analogue model of graphene is the graphene wormhole. In this setup, graphene sheets are arranged in such a way that they mimic some properties of wormholes in general relativity. There are several such models in the literature and here we address the graphene wormhole proposed by González and Herrero \cite{Gonzalez}. In this case, the wormhole is formed by a short carbon nanotube connecting two flat graphene sheets. The carbon nanotube acts as a bridge between the flat lattices and has a length much smaller than its radius. In addition, the junctions between the carbon nanotube and the flat sheets present six heptagonal rings alternating with hexagonal ones. González and Herrero found the emergence of zero modes when the effective gauge flux induced by topological defects reaches its maximum. Several studies have been carried out in this model since then, such as the behavior of fermions under the action of an axial magnetic field \cite{Rojjanason} and scattering properties \cite{Pimsamarn}. Additionally, the Casimir energy and tension \cite{Alencar}, the effects of curvature and deformation on the electronic properties \cite{silva2024strain}, and, recently, the quantum transport of massless fermions through the surface of the wormhole under the action of an axial magnetic field \cite{Naderi} were also studied. The existence of Landau levels for the graphene wormhole in the presence of an external magnetic field was also addressed in \cite{Garcia:2019gro}. More recently, Cavalcante {\it et al.} investigated the appearance of geometric phases for the graphene wormhole \cite{cavalcante2022quantum, cavalcante2024redshit}.
  
In this work, we focus on studying the behavior of fermion energy levels on rotating graphene wormholes. In order to construct the background geometry, we induce a rotational degree of freedom in the solution presented in \cite{Gonzalez} by direct analogy to the procedure usually performed on G\"odel spacetimes \cite{godel1949example}, which is the first solution of general relativity to admit a universe with a rotating perfect fluid. G\"odel-type geometries can be obtained simply from the transformation  $\theta'\rightarrow \theta + \omega t$ imposed on the angular coordinate of fixed background geometries, where $\omega$ denotes the angular velocity of rotated framework \cite{landau2013course}. It is known that on G\"odel-type geometries, Dirac fields can present Landau levels \cite{drukker2004godel, garcia2017fermions, garcia2018weyl} and recent works on this context involves the analysis of rotational effects on Dirac field in elastic media in the presence of dislocations \cite{maia2019topological, maia2020effects}. For graphene and its derivatives in particular, rotation effects have been addressed in the electronic spectrum of the fullerene molecule using an Abelian gauge field to describe topological defects and non-inertial effects \cite{lima2014effects, lima2015combined}. Moreover, the influence of rotation on the energy spectrum of carbon nanotube was investigated in \cite{cunha2015spin}. Recently, Brito discussed the role play of rotation in the quantum mechanics \cite{brito2024spinning}. Here, by dealing with the massless Dirac equation, we show that in the slow rotation limit the graphene wormhole behaves as an insulator material presenting landau levels analogue to G\"odel-type energy eigenvalues.  

This work is organized as follows: In Sec. \ref{sec2}, we recap some aspects of the graphene wormhole model described in \cite{Gonzalez}. In Sec. \ref{sec3}, we introduce rotational effects into the background geometry. In Sec. \ref{sec4}, we describe the Dirac equation for the rotating graphene wormhole, considering the metric both outside and inside the throat. In Sec. \ref{sec5}, we find the Landau levels for this configuration. Finally, in Sec. \ref{sec6}, we present the conclusions and ending comments.

%%%%%%%%%%%%%%%%%%%%%%%%%%%%%%%%%%%%%%%%%%%%%%%%%%%%%%%%%%%%%%%%%%%%%%%%%%%%%%
\section{Graphene wormhole}\label{sec2}

Graphene wormhole arises in the continuum limit of the model construct as follows. Two graphene sheets are connected by a bridge made of a carbon nanotube with zigzag boundaries. Such a bridge can be constructed by drilling a hexagonal-shaped hole in each graphene sheet, which makes it possible to connect the atoms on the hexagonal edges with new atoms in the vertical direction, creating a zigzag pattern with six heptagonal rings at each junction \cite{gonzalez2009propagating}. One considers setups where the radius of the hole is much larger than the nanotube length and, in this way, we can set aside soft curvature effects. Consequently, the background geometry curvature is zero everywhere except at the junction boundary. The emerging background geometry (for more details, see \cite{Gonzalez, Garcia:2019gro}) is given by the line element: 
\bes\label{splitmetric}\ben
ds^2&=&dr^2 + r^2 d\theta^2~~~,~~\text{for}~~ r\geq R,\\
ds^2&=&\left(\frac{R}{r}\right)^4\left(dr^2 + r^2 d\theta^2\right),~~\text{for}~~r\leq R,
\een\ees
where $R$ denotes the radius of the wormhole throat. The above metric might be expressed in terms of the conformal factor $A(r)$ as follows:
\begin{equation}
    ds^2=A(r)\left(dr^2 + r^2 d\theta^2\right)~~~,
\end{equation}
where 
\begin{equation}
    A(r)=\left(\frac{R}{r}\right)^4 \Theta(R-r)+\Theta(r-R),
\end{equation}
with $\Theta(x)$ is the well-known Heaviside step function. Now, we are interested in implementing spinning in the non-rotating wormhole metric and then exploring how rotating effects in the graphene wormhole affect the energy levels of the particles living on it. By doing so, we introduce a time-like coordinate $t$ by embedding the two-dimensional wormhole space into a $(2+1)$-dimensional spacetime. Therefore, the remaining metric reads
\begin{subequations}
\begin{eqnarray}\label{graphwormetric}
    ds^2 &=& -dt^2 + d\Omega^2,\ \ \text{for}\ r\geq R,\label{graphwormetricA}\\
    ds^2 &=& -dt^2 +\left(\frac{R}{r}\right)^4 d\Omega^2,\ \ \text{for}\ r < R. \label{graphwormetricB}
\end{eqnarray}
\end{subequations}
where $d\Omega^2=dr^2 + r^2 d\theta^2$.
%%%%%%%%%%%%%%%%%%%%%%%%%%%%%%%%%%%%%%%%%%%%%%%%%%%%%%%%%%%%%%%%%%%%%%%%%%%%%%
\section{Rotating graphene wormhole}\label{sec3}

A simple way to introduce rotation into the background geometry discussed in the previous section is by performing the transformation $\theta^{'} \rightarrow \theta + \omega t$  on the angular coordinate, which leads to a uniformly rotating wormhole with angular velocity $\omega$ around the radial coordinate \cite{landau2013course}. Hence, the line elements (\ref{graphwormetricA}) and (\ref{graphwormetricB}) becomes
\begin{subequations}
\begin{eqnarray}
    ds^2 &=& -\left(1 - r^2\omega^2\right)dt^2 + 2r^2 \omega\ dt d\theta+ d\Omega^2,~\text{for}\ r\geq R,\label{rotgraphA}\\
    ds^2 &=& -\left(1 - \frac{R^4}{r^2}\omega^2\right)dt^2 +\left(\frac{R}{r}\right)^{4}\left(2r^2 \omega\ dt d\theta + d\Omega^2\right),~\text{for}\ r < R, \label{rotgraphB}
\end{eqnarray}
\end{subequations}
describing graphene wormhole sectors outside $(r\geq R)$ and inside $(r < R)$ the throat, respectively. In the Cartan formalism, the ortonormal co-frame components $\hat{\theta}^{a} = e^{a}_{\ \mu}(x)\ dx^{\mu}$ are defined through the {\it vielbein} $e^{a}_{\ \mu}(x) e_{a}^{\ \mu}(x) = \delta^{a}_{\ b}$ considering that for locally flat spacetimes we have $ds^2 = \eta_{ab} \hat{\theta}^{a} \hat{\theta}^{b}$, where $\eta_{ab}$ represents the Minkowski tensor. Therefore, for $r\geq R$ we have
\begin{equation}
    e^{a}_{\ \mu}(x) = \left(\begin{array}{ccc}
    \sqrt{1 - r^2 \omega^2} & 0 & -\frac{r^2\omega}{\sqrt{1 - r^2 \omega^2}}\\
    0 & 1 & 0\\
    0 & 0 & \frac{r}{\sqrt{1 - r^2 \omega^2}} 
    \end{array}\right),~~~~~~~~
    e_{a}^{\ \mu}(x) = \left(\begin{array}{ccc}
    \frac{1}{\sqrt{1 - r^2 \omega^2}} & 0 & \frac{r\omega}{\sqrt{1 - r^2 \omega^2}}\\
    0 & 1 & 0\\
    0 & 0 & \frac{\sqrt{1 - r^2 \omega^2}}{r} 
    \end{array}\right),
\end{equation}
and for $r<R$ one finds
\begin{equation}
    e^{a}_{~\mu}(x) = \left(\begin{array}{ccc}
    \sqrt{1 - \frac{R^4}{r^2} \omega^2} & 0 & -\left(\frac{R}{r}\right)^2 \frac{R^2\omega}{\sqrt{1 - \frac{R^4}{r^2} \omega^2}}\\
    0 & \left(\frac{R}{r}\right)^2 & 0\\
    0 & 0 & \left(\frac{R}{r}\right)^2 \frac{r}{\sqrt{1 - \frac{R^4}{r^2} \omega^2}} 
    \end{array}\right),~~~~~~~~ 
    e_{a}^{~\mu}(x) = \left(\begin{array}{ccc}
    \frac{1}{\sqrt{1 - \frac{R^4}{r^2} \omega^2}} & 0 & \left(\frac{r}{R}\right)^2 \frac{(R^4\omega/r^3)}{\sqrt{1 - \frac{R^4}{r^2} \omega^2}} \\
    0 & \left(\frac{r}{R}\right)^2 & 0\\
    0 & 0 & \left(\frac{r}{R}\right)^2 \frac{\sqrt{1 - \frac{R^4}{r^2} \omega^2}}{r} 
    \end{array}\right). 
\end{equation}
The associated set of 2-fold spin connection components extracted from the torsionless Maurer-Cartan structures $d\hat{\theta}^a + \omega^{a}_{\ b}\wedge\hat{\theta}^b = 0$ is listed in the Table \eqref{tab1}, where $\omega_{ab}=\omega_{\mu ab}dx^\mu$ and $\omega_{ab}=-\omega_{ba}$. 
\begin{table}[h!]
\centering
\caption[]{Spin connection components}
\begin{tabular}{ccc}
\hline
\toprule
\begin{minipage}[t]{.25\textwidth} \centering {\bf Connection} \end{minipage}&
\begin{minipage}[t]{.25\textwidth} \centering {\bf $r\geq R$} \end{minipage}&
\begin{minipage}[t]{.25\textwidth} \centering {\bf $r < R$} \end{minipage}\\
\hline
%\toprule
\\[0.0005em]
$\omega^{\ 0}_{t\ 1} = ~\omega^{\ 1}_{t\ 0}$ & $-\frac{r\omega^2}{\sqrt{1-r^2 \omega^2}}$ & $\frac{R^2 \omega^2/r}{\sqrt{1-\frac{R^4}{r^2} \omega^2}} $	 \\[10pt]
%\hline
$\omega^{\ 1}_{t\ 2} = -\omega^{\ 2}_{t\ 1}$ & $-\frac{\omega}{\sqrt{1-r^2 \omega^2}}$ &$\frac{\omega}{\sqrt{1-\frac{R^4}{r^2} \omega^2}} $	 \\[10pt]
%\hline
$\omega^{\ 0}_{r\ 2} = ~\omega^{\ 2}_{r\ 0}$ & $\frac{\omega}{1-r^2 \omega^2}$ &$-\frac{R^2 \omega/r^2}{1-\frac{R^4}{r^2} \omega^2} $	 \\[10pt]
%\hline
$\omega^{\ 0}_{\theta\ 1} = ~\omega^{\ 1}_{\theta\ 0}$ & $-\frac{r\omega}{\sqrt{1-r^2 \omega^2}}$ &$-\frac{R^2\omega/r}{\sqrt{1-\frac{R^4}{r^2} \omega^2}}$	 \\[10pt]
%\hline
$\omega^{\ 1}_{\theta\ 2} = -\omega^{\ 2}_{\theta\ 1}$ & $-\frac{1}{\sqrt{1-r^2 \omega^2}}$ &$\frac{1}{\sqrt{1-\frac{R^4}{r^2} \omega^2}} $	 \\[10pt]
\hline
\toprule
\end{tabular}
\label{tab1}
\end{table}

%%%%%%%%%%%%%%%%%%%%%%%%%%%%%%%%%%%%%%%%%%%%%%%%%%%%%%%%%%%%%%%%%%%%%%%%%%%%%%
\section{Weyl equation for rotating graphene wormhole}\label{sec4}

As mentioned above, graphene presents a honeycomb lattice structure composed of carbon atoms. This lattice can be described by the superposition of two triangular sublattices $\mathcal{A/B}$ where each carbon atom is bounded with its first three neighbors. The valence electron is $\pi$-orbital and provides information about the electronic and band structures. The graphene tight-binding Hamiltonian can be expressed in terms of $\pi$-orbitals as \cite{katsnelson2007graphene}
\begin{eqnarray}
    H = -t\left(\sum_{\langle i,j\rangle} c^{\dagger}_{\mathcal{A},i} c_{\mathcal{B},j} + H.C.\right),
    \label{TBHamiltonian}
\end{eqnarray}
where $t$ denotes the hopping parameter, which gives us the probability of the valence electron hopping to a nearest neighbor. The tight-binding Hamiltonian (\ref{TBHamiltonian}) provides a dispersion relation where its first Brillouin zone has also a honeycomb lattice described by two Fermi points $K/K'$. In the low moment regime, the graphene band structure has a conical shape and its dispersion relation is linear, which implies that in this limit graphene behaves as a relativistic particle and we can use the massless Dirac equation to capture its energy levels. In this way, for graphene one can use the following Dirac equation,
\begin{eqnarray}
    H = -i\hslash v_f \Vec{\sigma}\cdot\Vec{k},
\end{eqnarray}
where $v_f$ represents the Fermi velocity. From here, we consider $v_f = \hslash =1$.  

The energy levels of graphene in curved geometries can be found from the Weyl equation
\begin{eqnarray}
    i\sigma^{\mu}(\nabla_{\mu} - iA_{\mu})\Psi = 0,   
    \label{weyleq}
\end{eqnarray}
where the covariant derivative operator $\nabla_{\mu} = \partial_{\mu} + \Gamma_{\mu}$ introduces the curved background elements, the spinorial connection is given by $\Gamma_\mu = \frac{1}{8} \omega_{\mu ab}[\sigma^a, \sigma^b]$ and $\lbrace\sigma^i\rbrace$ denotes the set of Pauli matrices in the real space. For the wormhole geometry presented in Eqs. \eqref{rotgraphA} and \eqref{rotgraphB} and using results from the Table \eqref{tab1}, one finds that the nonzero spinorial connections are 
\begin{eqnarray}
    \Gamma_t = \frac{i}{2} \omega^{\ 1}_{t\ 2} \sigma^3\ \ \text{and}\ \Gamma_\theta = \frac{i}{2} \omega^{\ 1}_{\theta\ 2} \sigma^3.
    \label{spinConec}
\end{eqnarray}
both inside and outside the wormhole throat.

In the graphene wormhole model proposed by Gonzalez and Herrero \cite{Gonzalez}, we have two graphene sheets with hexagonal holes connected by a zigzag carbon nanotube, which acts as a bridge between the graphene sheets. The atoms of the graphene sheet and the zigzag nanotube are bounded in the vertical direction - forming a graphene-nanotube junction - and due to this zigzag pattern, six heptagonal rings arise at each graphene-nanotube junction. The emergence of these heptagon rings in the graphene wormhole induces a negative curvature in the background geometry, which is associated with disclinations, causing electrons to hop to sites within the same sublattice and inducing a mixing of the $K/K'$ spinor components. In order to compensate this mixing, a $SU(2)$ gauge field can be introduced into Eq. \eqref{weyleq} as follows  \cite{pachos2009manifestations}:
\begin{eqnarray}
    A_{\theta} = \tau_2 \frac{\Phi}{2\pi},
\end{eqnarray}
here $\lbrace \tau_i\rbrace$ is the set of Pauli matrices in $K/K'$ space and $\Phi$ denotes a quantum flux. The gauge field $A_\theta$ represents topological defects in the junctions between the flat graphene sheets and the carbon nanotube. Moreover, we also have 
\begin{eqnarray}
    A_\theta = \pm\frac{\Phi}{2\pi},
    \label{gaugefield}
\end{eqnarray}
where $\pm 1$ are eigenvalues of the $\tau_2$ operator, representing the Abelian component of the projection in the $\tau$-space.  With all these ingredients, we can solve the Weyl equation \eqref{weyleq} for both the outer and inner sectors of the graphene wormhole throat. Here, we are investigating the massless Dirac equation in the low rotation regime, due {\color{red}} to the possibility of finding analytical solutions. In other words, we should consider $r\omega << 1$, or equivalently $\omega << 1$ for both cases, to obtain analytics solutions for the Dirac spinor.

\subsection{Outside of the throat ($r\geq R$)}

We start by dealing with the Weyl equation in the region outside the throat $(r\geq R)$. By using the spinorial connections \eqref{spinConec} together with the results presented in table \eqref{tab1} and introducing the gauge field \eqref{gaugefield} through a minimal coupling, we are led to the equation
\begin{eqnarray}\label{W1}
   i\sigma^0\left[\frac{\partial}{\partial t} + \frac{(\omega/2)\sigma^3}{\sqrt{1-r^2\omega^2}}\right]\Psi &+&i\sigma^1 \sqrt{1-r^2\omega^2} \left[D_{r} - \frac{r\omega^2}{2(1-r^2\omega^2)}\right]\Psi +i\sigma^2\frac{(1-r^2\omega^2)}{r} \left[D^{\pm}_{\theta} - \frac{r^2\omega}{\sqrt{1-r^2\omega^2}} \frac{\partial}{\partial t}\right]\Psi =0~~~~~~  
\end{eqnarray}
where, for simplicity, we define the auxiliary operators
\begin{equation} \label{Weyl1}
    D_{r}:=\frac{\partial~}{\partial r}+\frac{1}{2r}~~\text{and}~~D_{\theta}^{\pm}:=\frac{\partial~}{\partial \theta}\mp\frac{\Phi}{2\pi}.
\end{equation}
Moreover, in order to solve the Eq. \eqref{W1} we choose for the fermion field the {\it ansatz}
\begin{eqnarray}\label{a1}
    \Psi(t,r,\theta) = e^{-iEt + ij\theta} \left(\begin{array}{cc}
    \psi_{A}(r)\\
    \psi_{B}(r)
    \end{array}\right),
    \label{ansatz}
\end{eqnarray}
where each spinor component are representing the $\mathcal{A/B}$ sublattice wavefunction. By inserting the {\it ansatz} above into the Weyl equation \eqref{W1} we find a system of coupled first-order differential equations given by
\begin{widetext}
\begin{subequations}
\begin{eqnarray}
    \left[E + \frac{\omega/2}{\sqrt{1-r^2 \omega^2}}\right]\psi_{A} &=& -i\sqrt{1-r^2\omega^2}\left[D_{r} + \frac{r\omega^2/2}{1-r^2\omega^2}\right]\psi_{B} -i\left[\frac{1-r^2\omega^2}{r} \left(j \mp \frac{\Phi}{2\pi}\right) - r\omega E\right]\psi_{B},~~~~~~~~~\\
    \left[E - \frac{\omega/2}{\sqrt{1-r^2 \omega^2}}\right]\psi_{B} &=& -i\sqrt{1-r^2\omega^2}\left[D_{r} + \frac{r\omega^2/2}{1-r^2\omega^2}\right]\psi_{A} + i\left[\frac{1-r^2\omega^2}{r} \left(j \mp \frac{\Phi}{2\pi}\right) - r\omega E\right]\psi_{A},
\end{eqnarray}
\end{subequations}
\end{widetext}
which in the low rotation regime ($r\omega << 1$) can be reduced to the pair of equations
\begin{subequations}
\begin{eqnarray}\label{sr1}
    \left[E + \frac{\omega}{2}\right]\psi_{A} &=& -i\left[\frac{d}{dr} + \frac{1}{2r} + \frac{1}{r}\left(j\mp\frac{\Phi}{2\pi}\right) - r\omega E\right]\psi_{B},\\\label{sr2}
    \left[E - \frac{\omega}{2}\right]\psi_{B} &=& -i\left[\frac{d}{dr} + \frac{1}{2r} - \frac{1}{r}\left(j\mp\frac{\Phi}{2\pi}\right) + r\omega E\right]\psi_{A}.
\end{eqnarray}
\end{subequations}

\subsection{Inside of the throat ($r< R$)}

The procedure for dealing with the Weyl equation inside the wormhole throat is similar to that performed in the previous subsection. For this case, we have
\begin{eqnarray}\label{Weyl2}
   i\sigma^0\left[\frac{\partial}{\partial t} - \frac{(\omega/2)\sigma^3}{\sqrt{1-\frac{R^4}{r^2}\omega^2}}\right]\tilde{\Psi}&+&i\sigma^1 \left(\frac{r}{R}\right)^2 \sqrt{1-\frac{R^4}{r^2}\omega^2} \left[\tilde{D}_{r} - \frac{R^4\omega^2/r^3}{2\left(1-\frac{R^4}{r^2}\omega^2\right)}\right]\tilde{\Psi} +\\ \nonumber 
   &+&i\sigma^2 \left(\frac{r}{R}\right)^2 \frac{\left(1-\frac{R^4}{r^2}\omega^2\right)}{r} \left[\tilde{D}^{\pm}_{\theta} + \frac{R^4\omega/r^2}{\left(1-\frac{R^4}{r^2}\omega^2\right)} \frac{\partial}{\partial t}\right]\tilde{\Psi}=0,
\end{eqnarray}
where
\begin{equation}
    \tilde{D}_{r}:=\frac{\partial~}{\partial r}-\frac{1}{2r}~~\text{and}~~\tilde{D}_{\theta}^{\pm}:=\frac{\partial~}{\partial \theta}\mp\frac{\tilde{\Phi}}{2\pi},
\end{equation}
 and the {\it ansatz} we use now is
\begin{eqnarray}
    \tilde{\Psi}(t,r,\theta) = e^{-i\tilde{E}t + i\tilde{j}\theta} \left(\begin{array}{cc}
    \tilde{\psi}_{A}(r)\\
    \tilde{\psi}_{B}(r)
    \end{array}\right).
    \label{ansatz2}
\end{eqnarray}
By replacing this {\it ansatz} in the Weyl equation \eqref{Weyl2}, one finds a second system of coupled first-order differential equations, given by 
\begin{subequations}
\begin{eqnarray}
  \left[\tilde{E} \!-\! \frac{\omega/2}{\sqrt{1-\frac{R^4}{r^2} \omega^2}}\right]\!\tilde{\psi}_{A} &=& -i\left(\frac{r}{R}\right)^{\!2}\!\sqrt{1\!-\!\frac{R^4}{r^2}\omega^2}\left[\tilde{D}_r - \frac{R^4\omega^2/2r^3}{1-\frac{R^4}{r^2}\omega^2 } \right]\! \tilde{\psi}_{B}\! -\!i\left(\frac{r}{R}\right)^{\!2}\!\left[\frac{1\!-\!\frac{R^4}{r^2}\omega^2}{r} \left(\tilde{j} \mp \frac{\tilde{\Phi}}{2\pi}\right) \!-\! \frac{R^4\omega}{r^4}\right]\!\tilde{\psi}_{B},~~~~~~~~\\
 \left[\tilde{E} \!+\! \frac{\omega/2}{\sqrt{1-\frac{R^4}{r^2} \omega^2}}\right]\!\tilde{\psi}_{B} &=& -i\left(\frac{r}{R}\right)^{\!2}\!\sqrt{1\!-\!\frac{R^4}{r^2}\omega^2}\left[\tilde{D}_r - \frac{R^4\omega^2/2r^3}{1-\frac{R^4}{r^2}\omega^2}\right]\!\tilde{\psi}_{A} \!+\!i\left(\frac{r}{R}\right)^{\!2}\!\left[\frac{1\!-\!\frac{R^4}{r^2}\omega^2}{r} \left(\tilde{j} \mp \frac{\tilde{\Phi}}{2\pi}\right) \!-\! \frac{R^4\omega}{r^4}\right]\!\tilde{\psi}_A
\end{eqnarray}
\end{subequations}
and performing the coordinate transformation $\Tilde{r} = -R^2/r$ on the above equations, we have
\begin{widetext}
\begin{subequations}
\begin{eqnarray}
    \left[\tilde{E} - \frac{\omega/2}{\sqrt{1-\Tilde{r}^2 \omega^2}}\right]\tilde{\psi}_{A} &=& -i\sqrt{1-\Tilde{r}^2\omega^2}\left[\frac{\partial}{\partial \Tilde{r}} + \frac{1}{2\Tilde{r}} + \frac{\Tilde{r}\omega^2/2}{1 - \Tilde{r}^2\omega^2} \right]\tilde{\psi}_{B} +i\left[\frac{1-\Tilde{r}^2\omega^2}{\Tilde{r}} \left(\tilde{j} \mp \frac{\tilde{\Phi}}{2\pi}\right) - \Tilde{r}\omega \tilde{E}\right]\tilde{\psi}_{B};~~~~~~~\\
    \left[\tilde{E} + \frac{\omega/2}{\sqrt{1-\Tilde{r}^2 \omega^2}}\right]\tilde{\psi}_{B} &=& -i\sqrt{1-\Tilde{r}^2\omega^2}\left[\frac{\partial}{\partial \Tilde{r}} + \frac{1}{2\Tilde{r}} + \frac{\Tilde{r}\omega^2/2}{1-\Tilde{r}^2\omega^2}\right]\tilde{\psi}_{A} - i\left[\frac{1-\Tilde{r}^2\omega^2}{\Tilde{r}} \left(\tilde{j} \mp \frac{\tilde{\Phi}}{2\pi}\right) - \Tilde{r}\omega \tilde{E}\right]\tilde{\psi}_{A}.
\end{eqnarray}
\end{subequations}
\end{widetext}
In the low rotation limit where $\tilde{r}\omega << 1$, one finds the system 
\begin{subequations}
\begin{eqnarray}
    \left[\tilde{E} + \frac{\tilde{\omega}}{2}\right]\tilde{\psi}_{A} = -i\left[\frac{d}{d\Tilde{r}} + \frac{1}{2\Tilde{r}} - \frac{1}{\Tilde{r}}\left(\tilde{j}\mp\frac{\tilde{\Phi}}{2\pi}\right) -\Tilde{r}\tilde{\omega} \tilde{E}\right]\tilde{\psi}_{B},\\
    \left[\tilde{E} - \frac{\tilde{\omega}}{2}\right]\tilde{\psi}_{B} = -i\left[\frac{d}{d\Tilde{r}} + \frac{1}{2\Tilde{r}} + \frac{1}{\Tilde{r}}\left(\tilde{j}\mp\frac{\tilde{\Phi}}{2\pi}\right) + \Tilde{r}\tilde{\omega} \tilde{E}\right]\tilde{\psi}_{A},
\end{eqnarray}
\end{subequations}
where $\tilde{\omega}=-\omega$. Note that this set of equations has exactly the same structure of the set \eqref{sr1} and \eqref{sr2}.
%%%%%%%%%%%%%%%%%%%%%%%%%%%%%%%%%%%%%%%%%%%%%%%%%%%%%%%%%%%%%%%%%%%%%%%%%%%%%%
\section{Rotational Landau levels}\label{sec5}

Since when approaching the low rotation regime in the system under analysis we find similar equations for both the inner and outer sectors of the wormhole throat, we can treat them simultaneously here. In particular, by decoupling both pairs of equations we find two second-order differential equations for each case. In a compact way, one can write them as   
\begin{eqnarray}
  \frac{d^2 f_{\alpha,\sigma}}{dr_{\alpha}^2} + \frac{1}{r_\alpha}\frac{d f_{\alpha,\sigma}}{dr_\alpha} - \left\{ \omega_{\alpha} E_{\alpha} r_{\alpha}^{2} + \frac{\chi^{2}_{\alpha,\sigma}}{r_{\alpha}^{2}} - \beta_{\alpha,\sigma}\right\}f_{\alpha,\sigma} = 0, 
\end{eqnarray}
where $\alpha=1,2$ represents the regions outside and inside the throat, respectively, $(f_{1,\sigma},f_{2,\sigma}) = (\Psi,\tilde{\Psi})$, $(E_1,E_2)=(E,\tilde{E})$, $(r_1,r_2)=(r,\tilde{r})$, $(\Phi_1, \Phi_2)=(\Phi,\tilde{\Phi})$ and $(j_1, j_2)=(j,\tilde{j})$, $\sigma$ denotes the sublattice degrees of freedom and
\begin{subequations}
\begin{eqnarray}
    \chi^{2}_{\alpha,\sigma} &=& \left(j_{\alpha} \mp \frac{\Phi_{\alpha}}{2\pi} + \frac{\sigma_z}{2}\right)^2,\\
    \beta_{\alpha,\sigma} &=& E_{\alpha}^2 - \frac{\omega^2_{\alpha}}{4} + 2\omega_{\alpha} E_{\alpha}\left(j_{\alpha} \mp \frac{\Phi_{\alpha}}{2\pi} -  \frac{\sigma_z}{2} \right),
\end{eqnarray}
\end{subequations}
with $(\omega_1,\omega_2) = (\omega,\tilde{\omega})$ and $\sigma_z = 1, -1$ for $\mathcal{A, B}$ sublattices respectively. Performing the coordinate transformation $\xi_{\alpha} = \omega E_{\alpha} r_{\alpha}^2$, one finds the equation
\begin{eqnarray}
    \xi_{\alpha} \frac{d^2 f_{\alpha, \sigma}}{d\xi_{\alpha}^2} + \frac{df_{\alpha, \sigma}}{d\xi_{\alpha}} + \left[\frac{\beta_{\alpha, \sigma}}{4\omega_{\alpha} E_{\alpha}} - \frac{\xi_{\alpha}}{4} - \frac{\chi^{2}_{\alpha, \sigma}}{4\xi_{\alpha}}\right]f_{\alpha, \sigma} = 0,
    \label{secondEDO}
\end{eqnarray}
which has the asymptotic solution
\begin{eqnarray}
    f_{\alpha, \sigma}(\xi_{\alpha}) = e^{-\frac{\xi_{\alpha}}{2}} \xi_{\alpha}^{\frac{|\chi_{\alpha, \sigma}|}{2}} F_{\alpha, \sigma}(\xi_{\alpha}),
\end{eqnarray}
where $F_{\alpha, \sigma}(\xi_{\alpha}) =\ _{1}F_{1}\left(-\left[\frac{\beta_{\alpha, \sigma}}{4\omega E_{\alpha}} - \frac{1}{2} - \frac{|\chi_{\alpha, \sigma}|}{2}\right], |\chi_{\alpha, \sigma}|+1;\xi_{\alpha}\right)$ denotes the confluent hypergeometric function \cite{machado2012equaccoes}, which is solution of the confluent hypergeometric differential equation 
\begin{eqnarray}
    \xi_{\alpha}\frac{d^2 F_{\alpha, \sigma}}{d\xi_{\alpha}^2} + (|\chi_{\alpha, \sigma}| + 1 - \xi_{\alpha}) \frac{dF_{\alpha, \sigma}}{d\xi_{\alpha}} + \left(\frac{\beta_{\alpha, \sigma}}{4\omega E_{\alpha}} - \frac{1}{2} - \frac{|\chi_{\alpha, \sigma}|}{2}\right)F_{\alpha, \sigma} = 0,
\end{eqnarray}
By using the Frob\"enius method and truncating the solution of the above equation in a $n$-th order polynomial \cite{machado2012equaccoes}, one finds 
\begin{eqnarray}
    n_{\alpha} = \frac{\beta_{\alpha, \sigma}}{4\omega E_{\alpha}} - \frac{1}{2} - \frac{|\chi_{\alpha, \sigma}|}{2},
\end{eqnarray}
where $\left(n_1, n_2\right)=\left(n,\tilde{n}\right)$ are natural numbers used to describe the energy spectrum distribution of the field. Thus, the spinors on the rotating graphene wormhole background become
\begin{eqnarray}
    \Psi_{1}(t,\xi_1,\theta) = A_{n,j}\ e^{-iEt + ij\theta} e^{-\frac{\xi_1}{2}} \left(\begin{array}{cc}
    \xi_{1}^{|\chi_{1,\mathcal{A}}|} \ _{1}F_{1}\left(-n, |\chi_{1, \mathcal{A}}|+1;\xi_1\right)\\
     \xi_{1}^{|\chi_{1,\mathcal{B}}|} \ _{1}F_{1}\left(-n, |\chi_{1, \mathcal{B}}|+1;\xi_1\right)
    \end{array}\right)
    \label{wf1}
\end{eqnarray}
for $r\geq R$ and 
\begin{eqnarray}
    \Psi_{2}(t,\xi_2,\theta) = B_{\tilde{n},\tilde{j}}\ e^{-i\tilde{E}t + i\tilde{j}\theta} e^{-\frac{\xi_2}{2}} \left(\begin{array}{cc}
    \xi_{2}^{|\chi_{2,\mathcal{A}}|} \ _{1}F_{1}\left(-\tilde{n}, |\chi_{2, \mathcal{A}}|+1;\xi_2\right)\\
     \xi_{2}^{|\chi_{2,\mathcal{B}}|} \ _{1}F_{1}\left(-\tilde{n}, |\chi_{2, \mathcal{B}}|+1;\xi_2\right)
    \end{array}\right),
    \label{wf2}
\end{eqnarray}
for $r<R$, 
where $A_{n,j}$ and $B_{\tilde{n},\tilde{j}}$ denotes normalization constants to be determined.

It is noteworthy that the spinors \eqref{wf1} and \eqref{wf2} must obey the boundary condition where they are equals in the graphene wormhole junctions, i. e., we have that $\Psi_{1}(t, \omega E R^2, \theta) = \Psi_{2}(t, \omega \tilde{E} R^2, \theta)$. As one consequence, we have found relations among the spinors parameters:
\begin{equation}
    n=\tilde{n}, j = -\tilde{j},\ \Phi = -\tilde{\Phi},\ E = \tilde{E}\ \text{and}\ B_{n,-j} = e^{2ij\theta} A_{n,j}. 
\end{equation}
Therefore, the energy levels for the rotating graphene wormhole becomes
\begin{eqnarray}
    E_{n,j} &=& 2\omega\left[n + \frac{1}{2}\left| j \mp \frac{\Phi}{2\pi} \mp \frac{\sigma_z}{2}\right| - \frac{1}{2}\left(j \mp \frac{\Phi}{2\pi} \mp \frac{\sigma_z}{2}\right) + \frac{1}{2}(1 \pm \sigma_z)\right] \pm\nonumber\\   
    &~~~~&\pm 2\omega \sqrt{\left[n + \frac{1}{2}\left| j \mp \frac{\Phi}{2\pi} \mp \frac{\sigma_z}{2}\right| - \frac{1}{2}\left(j \mp \frac{\Phi}{2\pi} \mp \frac{\sigma_z}{2}\right) + \frac{1}{2}(1 \pm\sigma_z)\right]^2 + S^2}.
    \label{landau}
\end{eqnarray}

Note that, we have obtained the analogous Landau levels for rotated graphene wormhole in eq. \eqref{landau} in a compact way, representing both cases: $r\geq R$ and $r<R$. The energy eigenvalues possess a linear dependence with the angular velocity $\omega$ and the quantum flux given by $\Phi$. Here, $n$ is an integer number and $j_{\alpha}$ is a semi-integer number. It is important to point out that, in this case, the rotation effects are responsible for the rising of these bound states. Then, such effects mimic a magnetic field, as can be seen by a direct comparison with the result obtained in \cite{Garcia:2019gro}. We can observe that these energy levels presented in eq. \eqref{landau} is similar to energy eigenvalues for fermions for a class of G\"odel-type metrics \cite{garcia2017fermions}, specifically for fermions on the Som-Raychaudhuri spacetime with a cosmic string confined to a plane. The main difference between these two results is the break of degeneracy for each spinor component for the rotating graphene wormhole. We can see that the presence of the six heptagonal rings at the junction acts differently for each $K/K'$ valley. For both cases, the $\mathcal{A/B}$ sublattices are playing the role of spin so that the energy doublets emerge due to the interaction with the topological defects in the junctions. Another interesting point is the presence of the asymmetry term $S^2 = \frac{1}{16}$. This $S^2$ term is an energy gap term, and its addition is responsible for introducing the asymmetry for the energy levels for the rotating graphene wormhole. Notably, the $S^2$ term is analog to the torsion tensor for the landau levels for the Som-Raychaudhuri curved background obtained in ref. \cite{garcia2017fermions}. Differently of the graphene wormhole with an external magnetic field \cite{Garcia:2019gro}, the energy levels for rotated graphene wormhole \eqref{landau} is not purely Landau-type, so that the rotated graphene wormhole behaves with a pure insulator material without zero modes, due the $S^2$ asymmetry term.  

\begin{figure}[t!]
\centering 
\begin{tabular}{cc}    %%% not \center
\subfigure[The $\mathcal{A}$ sublattice and $K$ valley]{\includegraphics[width=70mm]{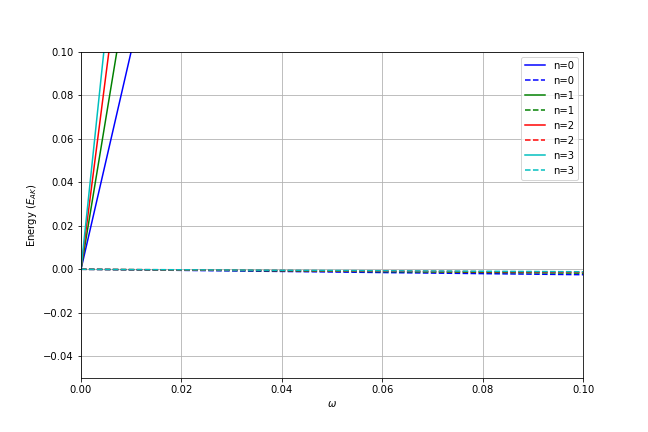}}
\subfigure[The $\mathcal{A}$ sublattice and $K'$ valley]{\includegraphics[width=70mm]{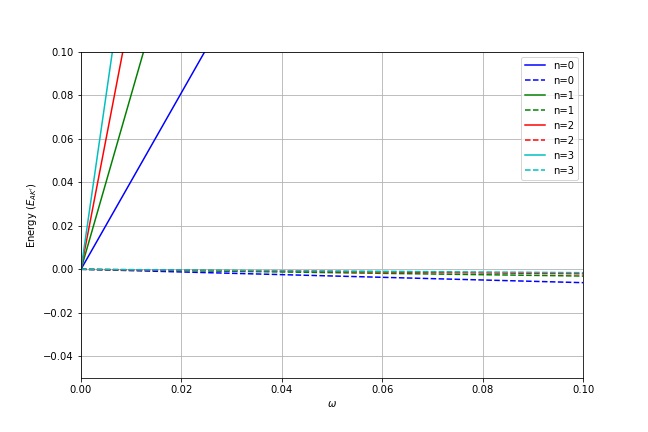}}
\end{tabular}
\begin{tabular}{cc}
\subfigure[The $\mathcal{B}$ sublattice and $K$ valley]{\includegraphics[width=70mm]{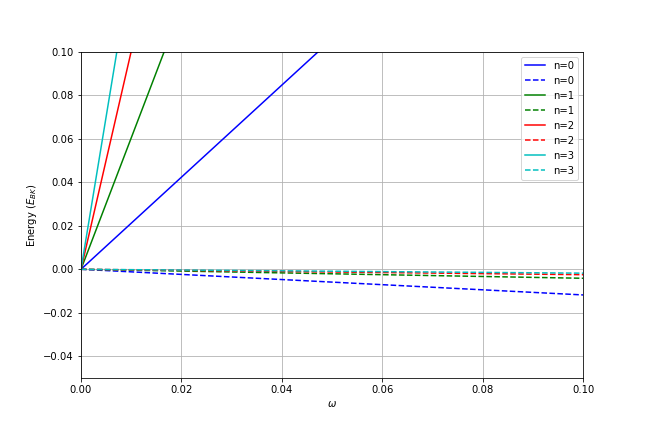}}
\subfigure[The $\mathcal{B}$ sublattice and $K'$ valley]{\includegraphics[width=70mm]{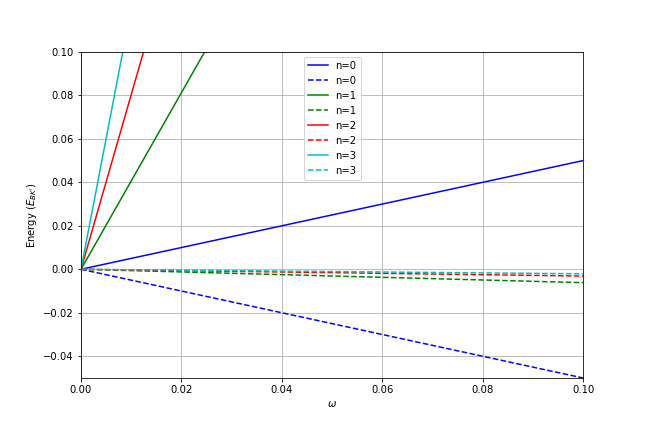}}
\end{tabular}
\caption{Energies levels for the rotating graphene wormhole.}\label{fig1}
\end{figure}

We depicted the energy levels in a set of two figures, namely: fig. \ref{fig1} is representing the energy levels for wormhole graphene for $r\geq R$ and $r<R$. For each graph, we chose in a range $n = 0, 1, 2, 3$, and the angular momentum quantum number was considered as $j = 1/2$. And, we have that the total quantum flux was admitted as $\Phi = 3\pi$. This choice was made considering the relative position of heptagon rings in the junctions. For this case, we considered that the distance between a pair of heptagons is a multiple of $3$. When this distance is not a multiple of 3, the total quantum flux is given by $\Phi = \pi$. We can see that for both cases (inside and outside of the throat), this set of figures \ref{fig1} showed to us the linear dependence with the angular velocity $\omega$ for the energy. The straight consequence of rotation effects is the degeneracy break of these energy levels for the sublattices and valleys degrees of freedom, where the valley degrees of freedom were introduced due to the presence of defects in the junctions. One concludes, from the assessment of the graphs, that the rotation affects the conductance and valence bands in a different way. It is clear the influence of the rotation in the conductance band, while the valence band exhibits a slight alteration in relation to zero energy. From figs. \ref{fig1} we also can observe the asymmetry of the landau levels introduced by the term $S^2$ and the presence of only one symmetric energy level when $n = 0$ for each case. These symmetric energy levels were previously zero modes, with their degeneracy broken by the rotation. Therefore we cannot observe any zero mode for the rotating graphene wormhole.

%%%%%%%%%%%%%%%%%%%%%%%%%%%%%%%%%%%%%%%%%%%%%%%%%%%%%%%%%%%%%%%%%%%%%%%%%%%%%%
\section{Summary and prospects}\label{sec6}

In summary, this work aimed to study the graphene wormhole as a gravitational analogue for a G\"odel-type spacetime, more specifically for the Som-Raychaudhuri curved background. With this purpose, we have described the graphene through a wormhole geometry model inspired by. \cite{Gonzalez}. This model connects two graphene sheets using a carbon nanotube as a bridge, at each junction between the nanotube and graphene sheets, six heptagon topological defects will appear. To simulate the Som-Raychaudhuri spacetime we carried out a coordinate transformation $\theta' \rightarrow \theta + \omega t$ in the graphene wormhole metric. Then, solving the Dirac equation for massless fermions in this rotating background, we have obtained Landau levels analogous to massless fermions in Som-Raychaudhuri spacetime \cite{garcia2017fermions} confined to plane. We have used the slow rotation approximation to obtain analytic solutions for the massless Dirac equation in the continuum limit for graphene. The energy levels found have a linear dependence on the angular velocity. The break of degeneracy entails the presence of topological defects in the junction and the effects of the rotation. The presence of topological defects breaks the degeneracy for the valley degrees of freedom, while the rotation effects break the degeneracy for the sublattice degrees of freedom. We still have shown that the rise of the $S^2$ term, in a similar way which occurs for fermions in Som-Raychaudhuri, is responsible for the asymmetry of the landau levels and is analog to the torsion tensor in \cite{garcia2017fermions}. Because of the existence of landau levels, it is possible to study the quantum Hall effect and the existence of persistent current in rotating graphene wormholes.    

%%%%%%%%%%%%%%%%%%%%%%%%%%%%%%%%%%%%%%%%%%%%%%%%%%%%%%%%%%%%%%%%%%%%%%%%%%%%%
{\bf Acknowledgements:} This work was supported by Conselho Nacional de Desenvolvimento Cient\'{\i}fico e Tecnol\'{o}gico (CNPq) and Funda\c{c}\~ao de Apoio a Pesquisa do Estado da Para\'iba (Fapesq-PB). G. Q. Garcia would like to thank to Fapesq-PB for financial support (Grant BLD-ADT-A2377/2024). P. J. Porf\'irio would like to thank the Brazilian agency CNPq for financial support (PQ--2 grant, process No. 307628/2022-1). The work by C. Furtado has been supported by the CNPq (project PQ Grant 1A No. 311781/2021-7).

\bibliographystyle{iopart-num}

  \bibliography{biblio} 

\end{document}